\documentclass[10pt,letterpaper]{article}
\usepackage{amsmath}
\usepackage{microtype}
\usepackage[utf8]{inputenc}
\usepackage{cite}
\usepackage{nameref,hyperref}
\usepackage[right]{lineno}
\usepackage{microtype}
\DisableLigatures[f]{encoding = *, family = * }
\raggedright
\usepackage{setspace} 
\usepackage{changepage}
\usepackage[aboveskip=1pt,labelfont=bf,labelsep=period,singlelinecheck=off]{caption}
\makeatletter
\renewcommand{\@biblabel}[1]{\quad#1.}
\makeatother
\usepackage[top=0.85in, left=1in, right=1in, footskip=0.75in, marginparwidth=2in]{geometry}
\usepackage{color}
\definecolor{Gray}{gray}{.25}
\usepackage{graphicx}
\usepackage{sidecap}
\usepackage[pscoord]{eso-pic}
\usepackage[fulladjust]{marginnote}
\reversemarginpar
\begin{document}
\begin{flushleft}
{\huge \textbf\newline{Case Study of Decoherence Times of Transmon Qubit} }
\newline
\newline
H. Zarrabi \textsuperscript{1},
S. Hajihosseini \textsuperscript{2},
M. Fardmanesh \textsuperscript{2},
S.I. Mirzaei \textsuperscript{1,*}
\\
\bigskip
\bf{1} Department of Condensed Matter Physics, Faculty of Basic Sciences, Tarbiat Modares University, Tehran, Iran  
\\
\bf{2} Department of Electrical Engineering, Sharif University of Technology, Tehran, Iran
\\
\bf{*} i.mirzaei@modares.ac.ir
\end{flushleft}
\section*{Abstract}
\justify
In the past two decades, one of the fascinating subjects in quantum physics has been quantum bits (qubits). Thanks to the superposition principle, the qubits can perform many calculations simultaneously, which will significantly increase the speed and capacity of the calculations. The time when a qubit lives in an excited state is called “Decoherence time.” The decoherence time varies considerably depending on the qubit type and materials. Today, short decoherence times are one of the bottlenecks in implementing quantum computers based on superconducting qubits. In this research, the topology of the transmon qubit is investigated, and the decoherence time caused by noise, flux, and critical current noise is calculated by numerical method.
\section*{Introduction}
\begin{figure}[htbp]
  \centering
  \includegraphics[width=65mm]{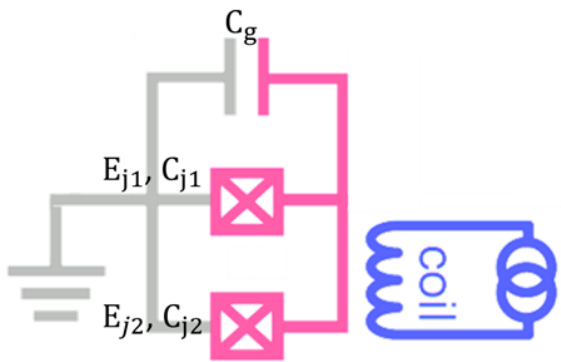}
  \caption{\centering Schematic of the transmon qubit \cite{[3]}}
  \label{fig1}
\end{figure}

To date, different qubit types have been introduced regarding working principles. One of the most popular of them is superconducting qubits. The Cooper pair box was the first charge qubit based on superconductivity.\cite{[1]} Due to its high sensitivity to charge noise, it was replaced by transmon qubit, which was introduced by Jens Koch et al. \cite{[2]} in 2008. The transmon qubit circuit consists of a SQUID (two parallel Josephson junctions) and a shunted capacitor, as shown in Figure \ref{fig1}. The parallel capacitor $(C_g)$ is much larger than the previous qubits. By adding a SQUID, which will reduce the coherence time by penetrating the flux noise in the loop, and increasing $E_J/Ec$, the key reason, the decoherence time could be improved by also incorporating an SQIUD and increasing $C_g$ in the circuit. Decoherence time is affected by the interaction of a qubit with the environment, which causes disturbances and collapse of superpositions and excitations to the ground state. Low decoherence time leads to errors in quantum information processing. The decay of the superposition of states is a limitation to performing complex quantum algorithms. On the other hand, there must be loose interactions between a qubit and its environment to read quantum data. Quantum decoherence is usually characterized by measuring two times: $T_1$ (relaxation time) and $T_2$ (dephasing time). Although in the new qubit, the sensitivity to charge noise is reduced, however; the flux sensitivity of the transmon qubit increases compared to the Cooper pair box qubit. 
\section*{Decoherence time}
Different parameters need to be considered, such as anharmonicity ($\alpha$) and decoherence time, when investigating the performance of the superconducting qubits. The time it takes for the coherence of the quantum state to be lost is called decoherence time. After decoherence time, the qubit can no longer continue its operation and it stops functioning correctly.
\subsection*{Dephasing time ($T_2$)}
Each state has an energy level, and the energy of a quantum transition between two energy levels is called transition energy. If environmental noises, such as charge noise, flux noise, or critical current noise, cause a change in the transition energy, dephasing occurs. When the noises coupled to the qubit and change the transition energy for the first two states of the system ($E_{01}$), dephasing time could be defined according to equation \ref{(1)} .\cite{[2]}
\begin{equation}
T_2 (\lambda) \approx \frac{\hbar}{A} \left| \frac{\partial E_{01}}{\partial \lambda} \right|^{-1}
\label{(1)}
\end{equation}
Where $\lambda$ could be a charge, flux, or critical current source noise,
$\hbar$ is the reduced Planck constant. $A$  is the amplitude of  $1⁄f$ noise and has different values depending on the type of noise, as mentioned in Table \ref{1}. The transition energies can be derived by utilizing the Hamiltonian of the transmon qubit.\cite{[2]}
\begin{equation}
\hat{H} = 4E_c (\hat{n} - n_g)^2 - E_{J\Sigma} \cos\left(\pi\phi_{\text{ext}}\right) \left[\cos\hat{\varphi} + d \tan\left(\pi\phi_{\text{ext}}\right) \sin\hat{\varphi}\right]
\label{(2)}
\end{equation}

Where ${\phi}/{\phi_0} = \phi_{\text{ext}}$, $E_{J\Sigma}$ is the sum of the two Josephson energies, which is dependent on the critical current, $E_c$ is the charging energy, which is dependent on the inverse of the total capacitance, and $d$ is the asymmetry coefficient of the Josephson junctions. Based on fabrication techniques, junction parameters could be different, with assumed junction asymmetries up to 10\%.\cite{[2]}
\begin{table}[!ht]
\centering
\begin{tabular}{|l|l|}
\hline
\multicolumn{1}{|c|}{\bf Dephasing} & \multicolumn{1}{|c|}{\bf Noise Source} \\
\hline
Charge & $A = 10^{-4}-10^{-3} \, e$ \\ \hline
Flux & $A = 10^{-6}-10^{-5} \, \Phi_0$ \\ \hline
Critical Current & $A = 10^{-7}-10^{-6} \, I_c$ \\ \hline
\end{tabular}
\label{1}
\vspace{7pt}
\caption{\centering Different values of the amplitude of $1/f$ noise based on charge, flux, and critical current noise.\cite{[2]}}
\label{1}
\end{table}
\newline
The second method of explaining the dephasing time is utilizing the Bloch sphere. If there is a coupling between environmental noises and the qubit along the $\hat{z}$ axis, dephasing occurs. The $\hat{z}$ axis represents the energy gap of the qubit. We expect an ideal qubit to have energy levels as constant as possible with respect to the noise $\lambda$. In Figure 2, you see the dephasing process, which includes the relaxation with a rate of $\Gamma_{1}/2$ and pure dephasing with a rate of $\Gamma_{\phi}$.\newline
\begin{figure}[htpb]
    \centering\includegraphics[width=90mm]{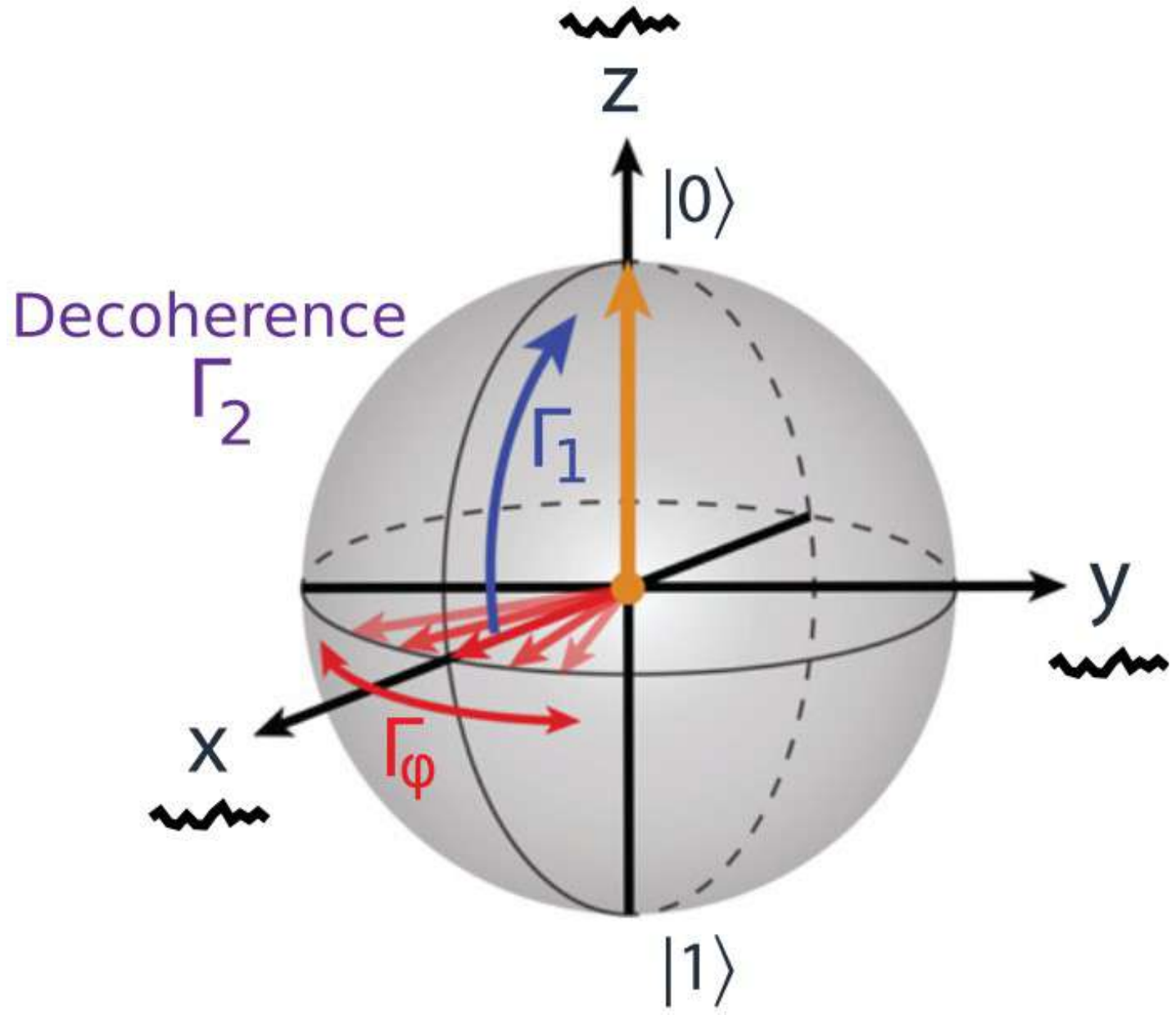}
    \newline
    \caption{\centering Dephasing time definition by Bloch sphere.\cite{[4]}}
  \label{fig2}
\end{figure}
\newline
In Figure 2, First of all, environmental noise affects the $\hat{z}$ axis and causes randomization of the phase of the qubit state, where the duration of this event is called the pure dephasing time ($T_{\phi}$). Then, by coupling between noise and the qubit along the two axes $\hat{x}$ and $\hat{y}$, it moves the state vector to a random position in the Bloch sphere. According to Figure 2, this phenomenon occurs at the rate of $\Gamma_{1}/2$, and the term $1/2T_1$ appears. In this way, equation \ref{(3)} describes the dephasing phenomenon with the help of the dephasing rate $\Gamma_2$.
\begin{equation}
\Gamma_2 = \frac{\Gamma_{1}}{2} + \Gamma_{\phi} = \frac{1}{T_2} = \frac{1}{2T_1} + \frac{1}{T_{\phi}} \label{(3)}
\end{equation}
Concerning equation \ref{(1)}, $d = 10\%$, $E_J = 20\, \text{GHz}$, and $E_c = 0.35\, \text{GHz}$ are assumed. The dephasing time of the transmon qubit caused by the charge, flux, and critical current noises calculated by Jens Koch et al., and the calculation values in the Quantum Toolbox in Python \cite{[5]} are shown in Table \ref{2}.\cite{[2]}
\newline
\vspace{7pt}
\begin{table*}[!ht]
\centering
\begin{tabular}{|c|c|c|}
\hline
\textbf{} & \textbf{Calculated by Jens Koch et al.} & \textbf{Calculation values in QuTiP} \\
\hline
Charge noise & $\approx 8$ $s$ & $8.667$ $s$ \\
\hline
Flux noise & $\approx 1$ $\mu s$ & $1.311$ $\mu s$ \\
\hline
Critical current noise & $\approx 35$ $\mu s$ & $32.104$ $\mu s$ \\
\hline
\end{tabular}
\caption{A comparison of the approximate values of the dephasing time ($T_2$) reported for the transmon qubit by Jens Koch et al. \cite{[2]} and values calculated in Python's Quantum Toolbox (QuTiP) \cite{[5]}.}
\label{2}
\end{table*}
\newline
In addition, with the help of Quantum Toolbox in Python and using equation \ref{(1)}, we can plot the diagram of dephasing times caused by all three noises based on different ${E_J}/{E_c} $, as shown in Figures \ref{fig3} and \ref{fig4}.
\newline
(a)\hspace{22em}(b)
\vspace{-45pt}
\begin{figure}[htpb]
    \centering\includegraphics[width=77mm]{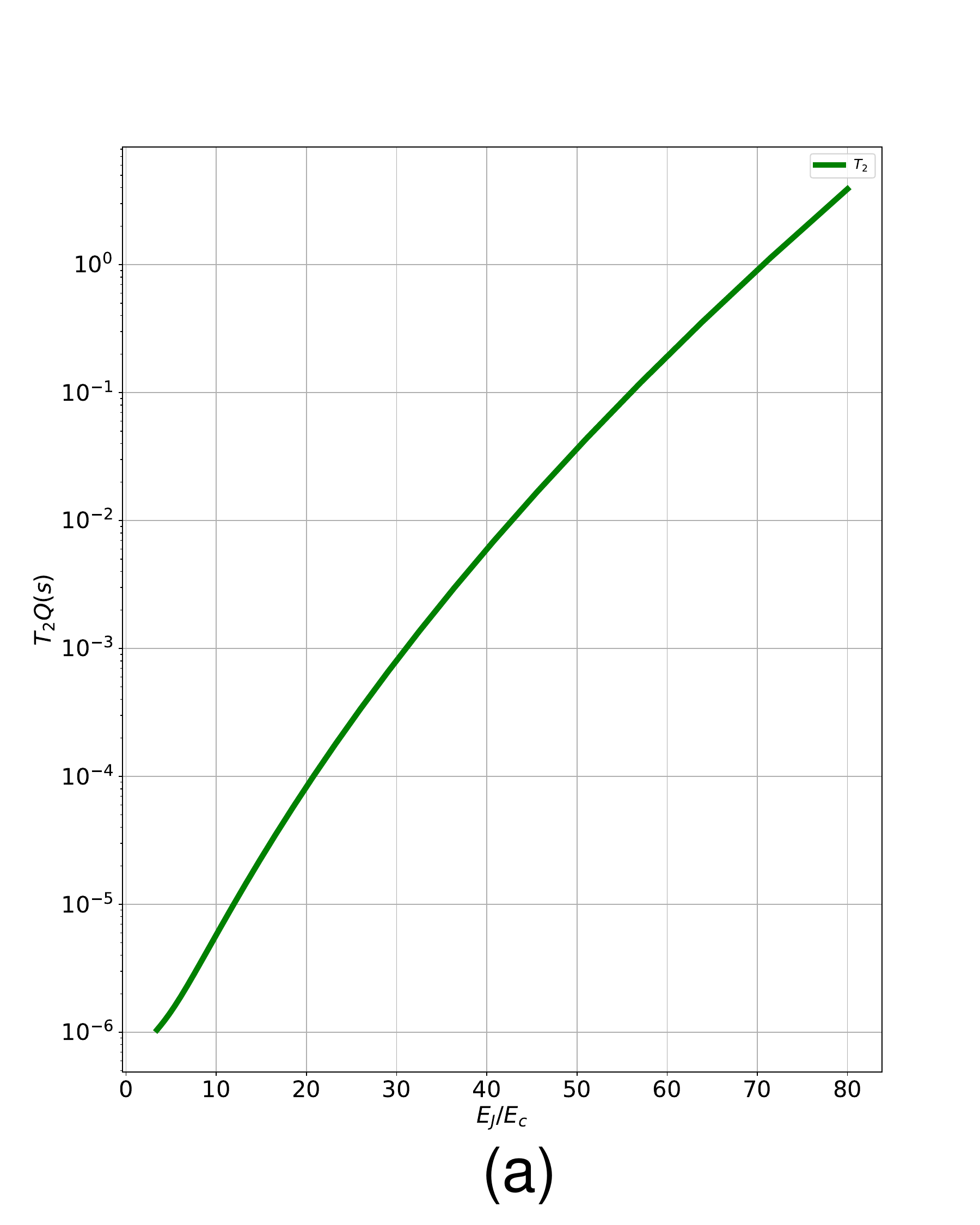}
    \centering\includegraphics[width=75mm]{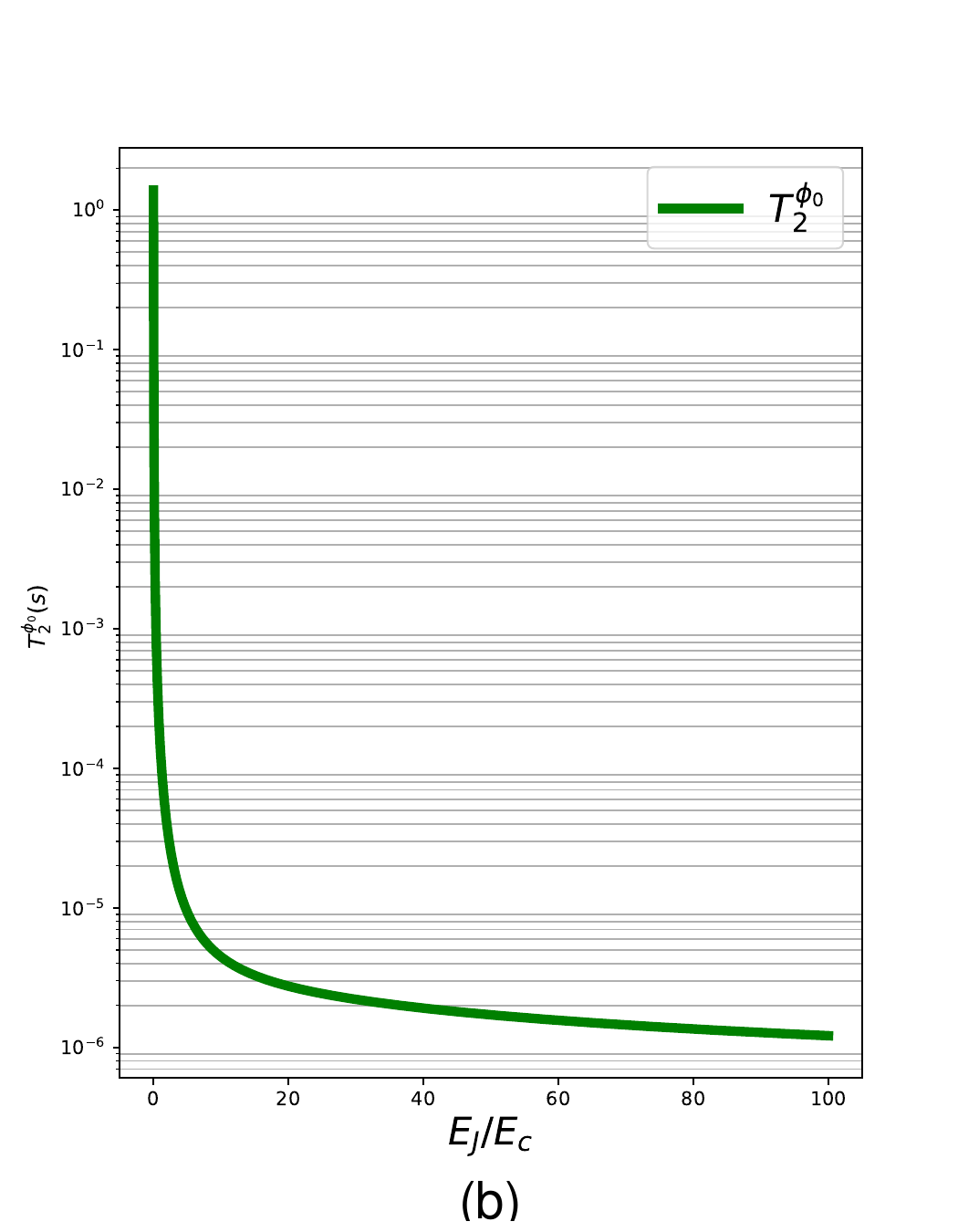}
    \newline
    \caption{\centering The diagram of dephasing time due to (a) charge noise ($T_2^Q$) and (b) flux noise ($T_2^{\phi}$) in terms of ${E_J}/{E_c}$.}
  \label{fig3}
\end{figure}
\newline
By comparing Figure 3(a) with the graph in ref.\cite{[2]}, one can verify the numerical results and see that the error percentage of this drawn graph on all  ${E_J}/{E_c}$ ratios is always less than 0.8\%.
\newpage
\vspace{-45pt}
(a)\hspace{19em}(b)
\vspace{-45pt}
\begin{figure}[htpb]
    \centering\includegraphics[width=73mm]{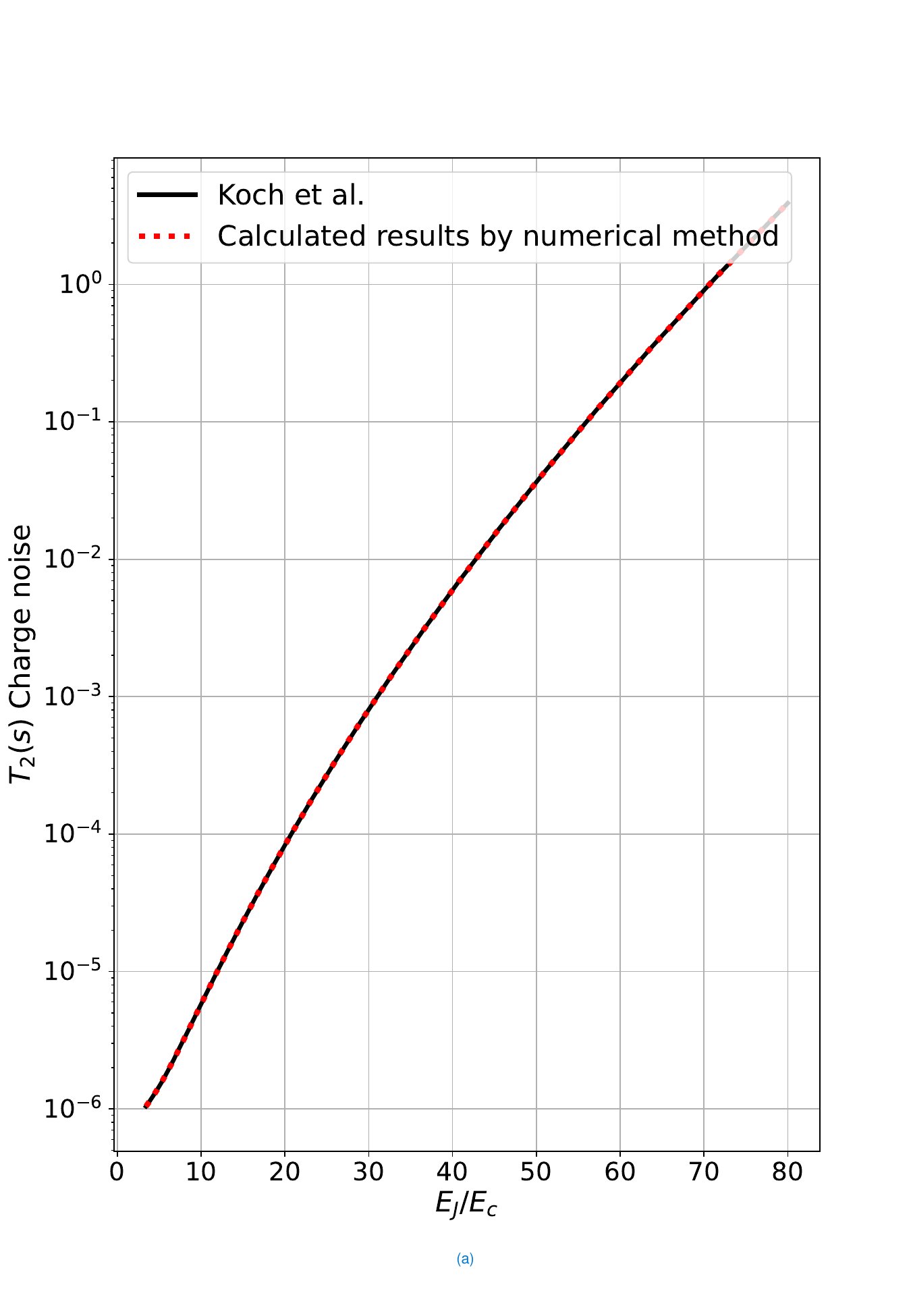}
    \centering\includegraphics[width=73mm, height=105mm]{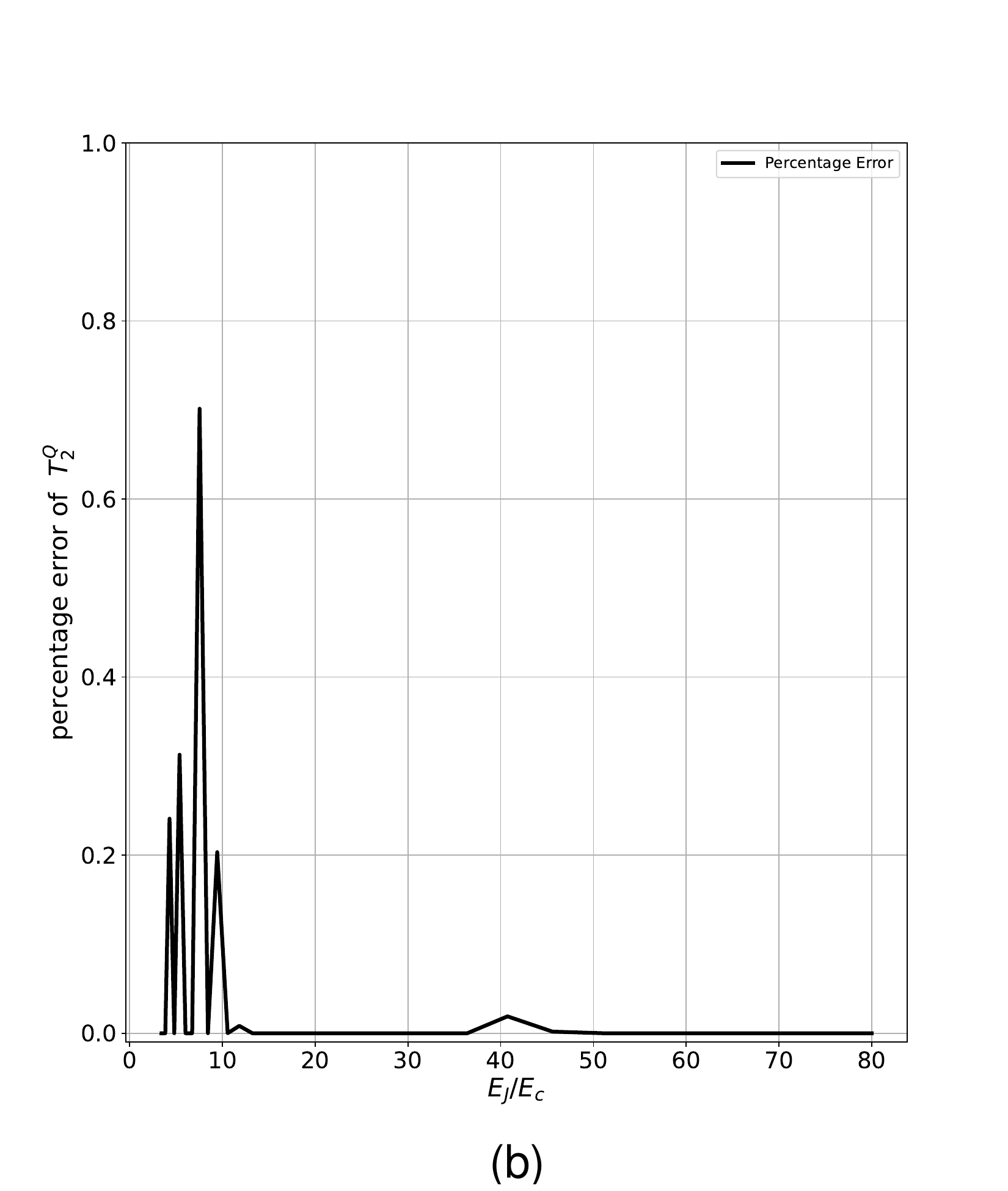}
    \newline
    \caption{\centering (a) Comparing the calculated results by numerical method with the results in ref \cite{[2]}. \newline (b) Error percentage in terms of ${E_J}/{E_c}$.}
  \label{fig4}
\end{figure}
\newline
In Figure 3, increasing the ${E_J}/{E_c}$ ratio, the dephasing time due to charge noise (\(T_2^Q\)) increases exponentially while the dephasing time due to flux noise (\(T_2^\phi\)) decreases. The reason for this behavior is evident in equations \ref{(4)} and \ref{(5)}. \cite{[2]}
\begin{equation}
    T_2^Q \propto e^{\sqrt{\frac{E_{J\Sigma}}{E_c}}} 
    \label{(4)}
\end{equation}
\begin{equation}
    T_2^\phi \propto \left(2E_c E_{J\Sigma}\right)^{-\frac{1}{2}} 
    \label{(5)}
\end{equation}
\newline
Regardless of the circuit topology, it is impossible to reduce the qubit sensitivity of charge and flux noise ratio simultaneously. This issue is one of the serious challenges nowadays.
\newpage
\begin{figure}[htbp]
\vspace{-30pt}
\centering 
    \begin{center}
        \includegraphics[width=95mm]{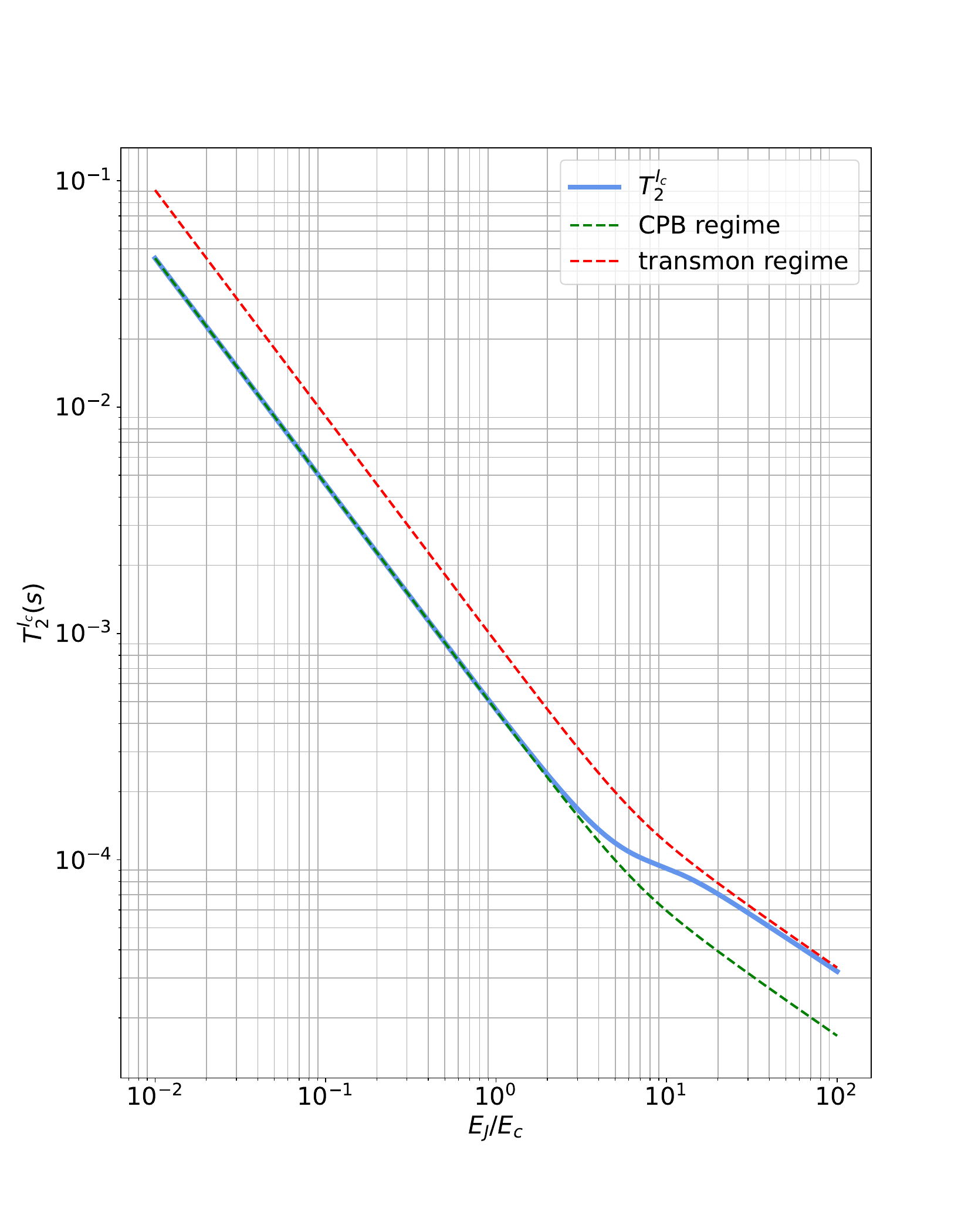}
        \newline
        \caption{\centering The diagram of dephasing time changes due to critical current noise in terms of ${E_J}/{E_c}$.}
        \label{fig5}
    \end{center}
\end{figure}
Another source of noise at low frequencies is Josephson's energy fluctuations. The source of critical current noise is the tunnel junction. Figure \ref{fig5} shows the changes in the duration of the dephasing due to the critical current noise with the increase of the ${E_J}/{E_c}$ ratio. According to equation \ref{(6)}, the dephasing time caused by the critical current noise is proportional to the inverse root of the critical current, which could be derived from equation \ref{(1)}. 
\begin{equation}
    T_2^{I_c} \propto \frac{\sqrt{I_c}}{I_c} = \frac{1}{\sqrt{I_c}} 
    \label{(6)}
\end{equation}
As a result, by increasing of \( I_c \) (or ${E_J}/{E_c}$ ratio), we expect the dephasing time caused by the critical current noise to decrease, as shown in Figure \ref{fig5}. The solid blue line in Figure \ref{fig5} is drawn based on the calculation results. The two red and green dotted lines are the result of the theoretical results of the Transmon and Cooper couple box regimes. \cite{[2]}
\section*{Conclusion}
Up to now, many attempts have been made to increase the dephasing and relaxation times by changing the topology of superconducting circuits. Optimizing the quantum circuits and preserving their valuable properties can create a new qubit with much better performance. To increase the dephasing time, it is necessary to increase all three dephasing times because, in reality, all the environmental noises are simultaneously coupled with the qubit.
\newline
Numerical simulations were conducted using Quantum Toolbox in Python \cite{[5]}to study the behavior of the $T_2$ dephasing time for various noise sources based on the $E_J/E_C$ ratio in both the Cooper pair box and transmon regimes. Employing equations from reference \cite{[2]}, the effects of different noise sources were examined. The simulation results demonstrated how the dephasing time responded to changing noise sources, with particular emphasis on critical current noise. The analysis suggests that an increase in $I_c$ (or $E_J/E_C$ ratio) may lead to a reduction in the dephasing time attributed to critical current noise. This trend aligns with theoretical expectations and highlights the intricate relationship between noise sources and qubit performance.
\vspace{35pt}
\bibliographystyle{plain}

\end{document}